# Understanding the Changing Roles of Scientific Publications via Citation Embeddings

Jiangen He[1]   Chaomei Chen[1]

[1] *{jiangen.he, chaomei.chen}@drexel.edu*
College of Computing and Informatics, Drexel University, Philadelphia (United States)

**Abstract**
Researchers may describe different aspects of past scientific publications in their publications and the descriptions may keep changing in the evolution of science. The diverse and changing descriptions (i.e., citation context) on a publication characterize the impact and contributions of the past publication. In this article, we aim to provide an approach to understanding the changing and complex roles of a publication characterized by its citation context. We described a method to represent the publications' dynamic roles in science community in different periods as a sequence of vectors by training temporal embedding models. The temporal representations can be used to quantify how much the roles of publications changed and interpret how they changed. Our study in the biomedical domain shows that our metric on the changes of publications' roles is stable over time at the population level but significantly distinguish individuals. We also show the interpretability of our methods by a concrete example.

**Conference Topic**
Indicators, Methods and techniques, Act of citations, in-text citations and Content Citation Analysis

**Introduction**
Although the content of a scientific publication cannot be changed once it was published, how other researchers cite and evaluate the publication may keep changing. The actual scientific contributions and impact of a specific publication are changing within the evolving intellectual spaces constructed with other publications. Besides the roles of publications are ever-changing, the role of a publication may be complex because of the varied contributions made by the publication, especially the publications that have contributed to interdisciplinary or fundamental research topics. The changing and complex roles of publications can be characterized by their citations.

We can construct citation networks and extract citation context (i.e., the sentences containing citations) for analyzing scientific publications (Elkiss et al., 2008). Citation networks are constructed commonly for characterizing a relevant intellectual structure and then identify the place where the analyzed publication is (Orosz, Farkas, & Pollner, 2016) or the alteration of the structure caused by the publication (Chen, 2012). The text of citation context is used to characterize publications for various applications, such as publication summarization (Qazvinian, 2010), survey article generation (Mohammad et al., 2009) and information retrieval (Huang, Wu, Liang, Mitra, & Giles, 2015). Quantitative metric for quantifying the role changes of publications can be derived from citation network analysis, but interpreting the changes is not straightforward which always relied on techniques of text mining and visual analytics; while analyzing the text of citation context naturally has interpretable results, a unified quantitative measurement is challenging to be built on the unstructured textual data.

In this article, we proposed a method to provide temporal representations of in-text citation of publications by word embedding models trained from citation context text, which can be used to characterize and analyze the changing and complex roles of the publications. The in-text citations of publications are the citations referred to this publication in the full text of other publications cited this publication; The text around the in-text citation is the citation context text. Based on the temporal representations of citations, we introduced a simple method to quantify the role changes of publications characterized by their citation context and described how to interpret the changes by making use of the interpretability features of embedding representations.

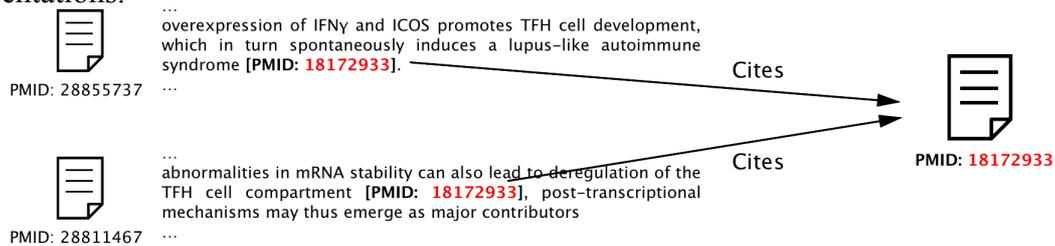

**Figure 1. Two examples of in-text citations of a PubMed article (PubMed ID: 18172933, "Roquin represses autoimmunity by limiting inducible T-cell co-stimulator messenger RNA."). The in-text citations of 18172933 are in the full text of articles who cited it.**

**Related work**

Recently, embedding learning techniques were employed in representing key elements of scientific knowledge, such as publications (Ganguly & Pudi, 2017), authors (Soumyajit, Manish, Varma, & Pudi, 2015), citations (Berger, McDonough, & Seversky, 2017) and research topics (He & Chen, 2017). Paper2vec (Ganguly & Pudi, 2017) leveraged both citation networks and textual information of publications to represent a publication, but the textual information they used is the full text of publications which is the description from authors of publication rather than scientific communities. Another study also named 'Paper2vec' (Tian & Zhuo, 2017) focused on utilizing neighbor nodes of publications in citation network to represent the publications. Cite2vec (Berger et al., 2017) represented publications by using their citation context as ours and provided a visualization for exploration, while the temporal feature of citation context is ignored. In our study, we emphasize representing the changes of publications characterized by the citation context of the publications over time.

**Methods**

In this section, we describe how we train temporal citation embedding models, which includes embedding model training for each period and embedding model alignment. We also present our approaches to quantifying the role changes of publications and interpreting the changes based on the trained temporal embedding models.

*Data and preprocessing*

The dataset we used for training is the PubMed Central Open Access Subset (PMC OAS) which is an open access XML formatted full-text document repository from biomedicine and the life sciences maintained by the U.S. National Library of Medicine (NLM)[1]. We can parse in-text citations and citation context from PMC OAS because of the well-structured XML files. In this study, we trained embedding models by documents published from 2007 to 2016 in PMC OAS which comprises 1,197,931 full-text scientific publications.

To train the citation embeddings, we need to use a unique identifier to indicate a publication in full text. Many references cited by publications in the PMC OAS have unique publication identifiers such as PubMed ID (PMID) and PubMed Central ID (PMCID). However, many cited references are not included in PubMed, so they don't have unique identifiers. We assign unique identifiers for these references by using their metadata in the form of 'FA_VE_YR_VO_FP' where FA is the first author's first name and last name, VE is the name of venue (journal or conference proceeding), YR is the year of the publication date, VO is the volume number, and FP is the first page number of the publication. If a cited reference has neither a PubMed ID nor identifiable metadata for constructing a unique identifier, the reference will be ignored in this study.

---

[1] http://www.ncbi.nlm.nih.gov/pmc/

Since our embedding learning method learns the representation of a citation by capturing the context of the citation within its sentence, we need sentence tokenizer to segment the full text into sentences. Then, we conduct a series of preprocessing by using 'NLPre'[2], including dash removal, token replacement, URL replacement, capitalization normalization, replacing phrase from Medical Subject Headings (MeSH) dictionary, etc.

*Embedding learning*

Word embedding techniques were proved to be able to capture semantic and syntactic effectively (Mikolov, Sutskever, Chen, Corrado, & Dean, 2013). We build citation embeddings for understanding how researchers described the cited publication. We use skip-gram with negative sampling (SGNS) (Mikolov et al., 2013) to learn citation embedding based on the context words of citation in the full text of publications. Given a citation or a phrase $w_i$ in training dataset, skip-gram maps it into a continuous representation vector $\mathbf{w}_i$. $\mathbf{w}_i$ is used to predict the context words of $w_i$. The objective of skip-gram is to maximize the log probability:

$$\frac{1}{T}\sum_{i=1}^{T}\sum_{i-c \leq j \leq i+c} \log p(w_j|w_i)$$

where $T$ is the occurrence of each word or citation in the training data, $c$ is the window size of context and $w_j$ is the context of $w_i$. Negative sampling builds 'negative' context words for each $w_i$ to accelerate the training procedure. We leave out the technical details of skip-gram and negative sampling because of limited pages. We separately constructed citation embeddings from publication text data for each year by SGNS algorithm. We empirically set embedding length as 100 and the window size as 5.

*Temporal embedding alignment*

Since our embedding models are constructed separately for different years, the models are in different vector space. We need to align the models for different years into the same coordinate axes to quantify and compare the citation context changes of a publication. We use orthogonal Procrustes to align the learned low-dimensional embeddings (Hamilton, Leskovec, & Jurafsky, 2016). Defining $\mathbf{W}^{(t)} \in \mathbb{R}^{d \times |v|}$ as the matrix of word embeddings learn at period $t$, we align across time periods while preserving cosine similarities by optimizing

$$\mathbf{R}^{(t)} = \arg\min_{\mathbf{Q}^T\mathbf{Q}=\mathbf{I}} \left\| \mathbf{Q}\mathbf{W}^{(t)} - \mathbf{W}^{(t+1)} \right\|_F$$

with $\mathbf{R}^{(t)} \in \mathbb{R}^{d \times |v|}$.

*Quantify the changes of citation context*

The representations of citations can be compared over different years after the embedding models derived from different years were aligned. The difference between representations of a cited publication in different years can be utilized to quantify the changes of the publication's citation context. We measure the difference based on commonly used cosine similarity. Therefore, we quantify the changes of publication $p$ happened at year $t$ as

$$Change_p^t = 1 - \cos\_sim(\mathbf{w}_p^t, \mathbf{w}_p^{t-1})$$

where $\mathbf{w}_p^t$ is the vector representation of publication $p$ at year $t$ derived from the citation context of $p$ at $t$.

---

[2] https://pypi.python.org/pypi/nlpre, a natural language pre-processor.

*Understanding the changing roles of publications*

The temporal representations of publications have the interpretability of the changing roles of publications because the representations can be described by other related and readable words or phrases in the embedding models. The related words can be easily identified by finding the nearest neighbours of the publication representation in the vector space. By identifying the related word to a publication over years, we may gain a picture depicting how the role of the publication changed in the scientific community within these years. We showed a concrete example in the section of Result.

**Results**

*Data descriptions*

We used 1,361,455 full-text scientific publications from PMC OAS in a recent decade for our analysis, but the publications without a citation of identifiable publications in the full text will be excluded. 1,205,407 publications have at least one effective citing sentence and they have 31 citing sentences on average. About 37.5 million sentences were used for our embedding model training. In this study, we aim to represent cited publications by their citation context, so we focus on the cited publications (CP) which have enough citation context information for representations. We identified cited publications with more 20 in-text citations for further analysis. We show the data distribution in Table 1.

**Table 1. Distribution of publications and cited publications in PMC OAS from 2007 to 2016.**

| Year | Publications | Publications with citations | CP | CP with in-text citations > 20 |
|---|---|---|---|---|
| 2007 | 27,062 | 19,691 | 499,612 (427,987) | 1,602 (1,553) |
| 2008 | 42,332 | 29,414 | 721,753 (615,072) | 2,451 (2,369) |
| 2009 | 58,531 | 40,014 | 950,542 (727,542) | 3,697 (3,590) |
| 2010 | 79,673 | 52,759 | 1,270,089 (1,038,389) | 5,733 (5,544) |
| 2011 | 108,461 | 95,274 | 2,019,401 (1,593,430) | 11,567 (11,176) |
| 2012 | 144,253 | 129,827 | 2,676,729 (2,103,427) | 19,512 (18,961) |
| 2013 | 182,261 | 165,410 | 3,284,862 (2,532,453) | 26,772 (25,985) |
| 2014 | 216,359 | 195,492 | 3,827,767 (2,907,255) | 32,282 (31,207) |
| 2015 | 244,293 | 229,620 | 4,478,018 (3,335,289) | 40,653 (39,156) |
| 2016 | 258,230 | 247,906 | 4,896,693 (3,586,434) | 46,182 (44,513) |

[1] The number in the parentheses is the number of CPs included in PubMed.

*Change Scores*

We computed the change score for each CP in each year by the method we introduced and observed the temporal distribution of the scores by three groups (see Table 2). Since the change score of $p$ at $t$ is derived from its representation at $t$ and $t$-1, we observe the distribution from 2008 to 2016. Table 2 shows that the change scores are stable over time within a group, which are barely affected by the factors of time and observation number. However, the inter-group difference is obvious, which may indicate that the more frequently a CP is cited, the more stable its role tends to be.

We also analyzed change score value distribution of the three groups (see Figure 2). The three groups have similar distributions where most of CPs' change scores lie in the range of 0.05 to 0.2, but the groups with more citations have narrower score range. Additionally, the distributions are roughly normal, but the groups with more citations are less slightly peaked than the ones with fewer citations.

Table 2. Temporal distribution of the change scores of CPs from 2008 to 2016[1].

| Year | CP with citations > 20 | CP with citations > 50 | CP with citations >100 |
|---|---|---|---|
| 2008 | 0.107 (*SD*=0.035, *N*=736) | 0.092 (*SD*=0.025, *N*=114) | 0.084 (*SD*=0.016, *N*=28) |
| 2009 | 0.110 (*SD*=0.039, *N*=1,213) | 0.092 (*SD*=0.026, *N*=197) | 0.085 (*SD*=0.019, *N*=49) |
| 2010 | 0.110 (*SD*=0.037, *N*=1,836) | 0.094 (*SD*=0.026, *N*=293) | 0.086 (*SD*=0.023, *N*=71) |
| 2011 | 0.113 (*SD*=0.036, *N*=3,172) | 0.097 (*SD*=0.028, *N*=482) | 0.086 (*SD*=0.022, *N*=116) |
| 2012 | 0.113 (*SD*=0.037, *N*=6,477) | 0.096 (*SD*=0.029, *N*=997) | 0.086 (*SD*=0.028, *N*=233) |
| 2013 | 0.113 (*SD*=0.038, *N*=10,719) | 0.094 (*SD*=0.031, *N*=1,688) | 0.085 (*SD*=0.038, *N*=420) |
| 2014 | 0.112 (*SD*=0.038, *N*=14,344) | 0.093 (*SD*=0.028, *N*=2,415) | 0.086 (*SD*=0.031, *N*=616) |
| 2015 | 0.113 (*SD*=0.038, *N*=17,670) | 0.095 (*SD*=0.029, *N*=3,103) | 0.086 (*SD*=0.032, *N*=802) |
| 2016 | 0.112 (*SD*=0.038, *N*=21,859) | 0.093 (*SD*=0.026, *N*=3,926) | 0.082 (*SD*=0.022, *N*=1,054) |

[1] *SD* = Standard deviation, *N* = the Number of observed CPs.

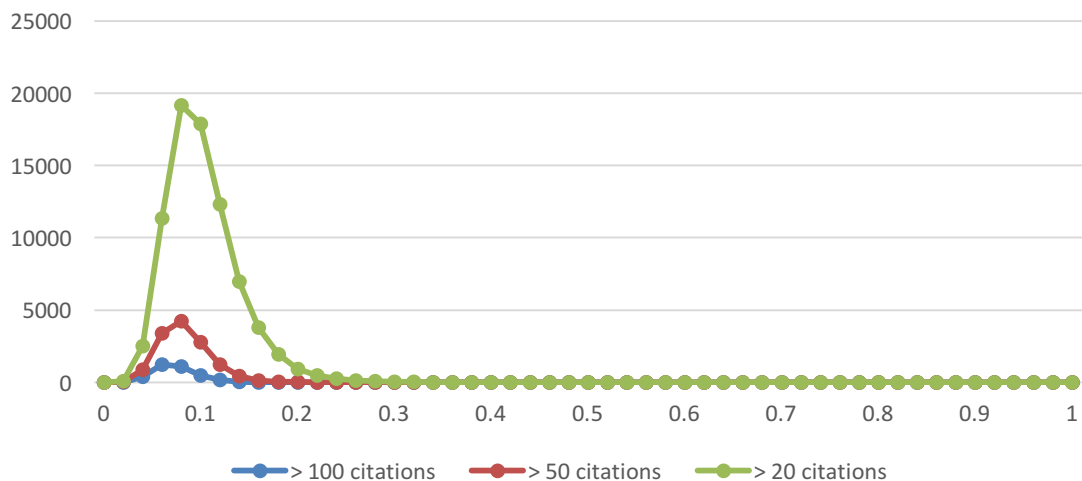

Figure 2. The distribution of change score value.

*Change score ranking*

We listed 5 publications with highest average change scores over recent 5 years (2012-2016). These publications greatly changed descriptions from other researchers in recent five years. The high change scores may be indications of various reasons, such as high novelty or controversy. The underlying reasons need a further examination.

Table 3. Top 5 publications with highest average change score over recent 5 years (2012-2016)

| PMID | Published year | In-text citations | Actual citations[1] | Avg. change score |
|---|---|---|---|---|
| 17081983 | 2006 | 392 | 2,804 | 0.186 |
| 12845331 | 2003 | 423 | 1,656 | 0.156 |
| 17081983 | 2005 | 379 | 2,194 | 0.151 |
| 18171944 | 2008 | 594 | 1,684 | 0.143 |
| 19372393 | 2009 | 375 | 1,711 | 0.141 |

[1] from Google Scholar.

*Interpretability*

The change scores alone are not informative for us to understand the changes of citation context of a publication. Based on the citation representations, we can retrieve a series of similar words and publications at each time to interpret the changes. We use the publication with the highest average change score over recent 5 years (Olsen et al., 2006) as an example to demonstrate the interpretability of the temporal representations of publications. We listed the most similar publication and a group of most similar words of the publication at each year from 2012 to 2016 in Table 4. From Table 4, we can see the words which describe the original content of the publication like 'phosphosite'; we can also see the words describing the scientific development related to this publication like 'kinase specific phosphorylation site prediction' and 'UbPred'. It is worth noting that most similar items to publications are other publications rather than words. It is quite reasonable because publications naturally share more syntactic and semantic features than with words. The similar publications may also provide a proxy to understand the changes.

Table 4. The changes of citation context of PubMed article 17081983, "Global, in vivo, and site-specific phosphorylation dynamics in signaling networks", 2006, Cell.

| Year | Change Score | The most similar publications[1,2] | Most similar words[2] |
|---|---|---|---|
| 2012 | 0.168 | 20068231 (0.84) | PHOSIDA (0.78), PhosPhAt (0.77), PhosphoSite (0.76), MiCroKit (0.75), NetPhos (0.74), ChloroP1.1 (0.74), CisGenome (0.74), Phospho.ELM (0.74) |
| 2013 | 0.215 | 21177495 (0.79) | Guittard (0.74), Scansite (0.70), phosphosite, (0.70), Tyr216 (0.69), pY (0.68), AKT (0.68), Sarbassov (0.68), IRAG (0.68), phospho-protein (0.68) |
| 2014 | 0.198 | 21183079 (0.83) | phosphopeptide (0.73), Phosida (0.73), NetPhos (0.73), ChIP-seq (0.73), SignalP4.1 (0.72), Scansite (0.72), kinase specific phosphorylation site prediction (0.72), mNgn2 (0.71) |
| 2015 | 0.145 | 20068231 (0.77) | phosphosite (0.75), phosphopantetheinylation (0.71), OGlcNAcylation, (0.71), Schwanhausser (0.70), phosphotyrosine-containing (0.70), phosphopeptide (0.69), phosphoamino (0.69), PTM (0.69) |
| 2016 | 0.201 | 21081558 (0.84) | Hornbeck (0.80), UbPred (0.79), PHOSIDA (0.79), NetPhosK (0.78), PhosphoSitePlus (0.78), NetPhos (0.76), PhosphoSite (0.76), pY (0.75) |

[1] PMID. [2] The value in the parentheses is similarity score.

**Conclusions**

In this article, we introduced an embedding learning method to represent scientific publications by using the citation context text in other publications. Our method emphasizes the temporal features of citation text to characterize the dynamic role of scientific publications. The temporal representation can be used to quantify how much the role of a publication changed as well as interpret how the role changed over time. Base on the study on a large biomedical full-text literature dataset, we conclude that the metric for quantifying the changes of publications' roles is stable over time at the population level and there is significant individual variability to distinguish individuals.

**Acknowledgments**

The work is supported by the National Science Foundation (Award Number: 1633286). We thank Zhipeng Zheng at Department of Chemistry, University of Pennsylvania for his help in the example interpretation.

# References

Berger, M., McDonough, K., & Seversky, L. M. (2017). Cite2vec: Citation-Driven Document Exploration via Word Embeddings. *IEEE Transactions on Visualization and Computer Graphics*, *23*(1), 691–700. http://doi.org/10.1109/TVCG.2016.2598667

Chen, C. (2012). Predictive effects of structural variation on citation counts. *Journal of the American Society for Information Science and Technology*, *63*(3), 431–449.

Elkiss, A., Shen, S., Fader, A., Erkan, G., States, D., & Radev, D. (2008). Blind men and elephants: What do citation summaries tell us about a research article? *Journal of the Association for Information Science and Technology*, *59*(1), 51–62.

Ganguly, S., & Pudi, V. (2017). Paper2vec: Combining Graph and Text Information for Scientific Paper Representation. In *European Conference on Information Retrieval* (pp. 383–395).

Hamilton, W. L., Leskovec, J., & Jurafsky, D. (2016). Diachronic Word Embeddings Reveal Statistical Laws of Semantic Change. Retrieved from http://arxiv.org/abs/1605.09096

He, J., & Chen, C. (2017). Predictive Effects of Novelty Measured by Temporal Embeddings on Growth in Science. *Manuscript Submitted for Publication*.

Huang, W., Wu, Z., Liang, C., Mitra, P., & Giles, C. L. (2015). A Neural Probabilistic Model for Context Based Citation Recommendation. In *AAAI 2015: Proceedings of the Twenty-ninth AAAI Conference on Artificial Intelligence* (pp. 2404–2410).

Mikolov, T., Sutskever, I., Chen, K., Corrado, G., & Dean, J. (2013). Distributed Representations of Words and Phrases and Their Compositionality. In *Proceedings of the 26th International Conference on Neural Information Processing Systems* (pp. 3111–3119). USA: Curran Associates Inc. Retrieved from http://dl.acm.org/citation.cfm?id=2999792.2999959

Mohammad, S., Dorr, B., Egan, M., Hassan, A., Muthukrishan, P., Qazvinian, V., … Zajic, D. (2009). Using citations to generate surveys of scientific paradigms. In *Proceedings of Human Language Technologies: The 2009 Annual Conference of the North American Chapter of the Association for Computational Linguistics* (pp. 584–592).

Olsen, J. V, Blagoev, B., Gnad, F., Macek, B., Kumar, C., Mortensen, P., & Mann, M. (2006). Global, In Vivo, and Site-Specific Phosphorylation Dynamics in Signaling Networks. *Cell*, *127*(3), 635–648. http://doi.org/10.1016/j.cell.2006.09.026

Orosz, K., Farkas, I. J., & Pollner, P. (2016). Quantifying the changing role of past publications. *Scientometrics*, *108*(2), 829–853.

Qazvinian, V. (2010). Citation Summarization Through Keyphrase Extraction. *Computational Linguistics*, (August), 895–903. Retrieved from http://www.aclweb.org/anthology/C10-1101

Soumyajit, G. J., Manish, G., Varma, V., & Pudi, V. (2015). Author2Vec: Learning author representations by combining content and link information. *Acm*, (3), 4503. http://doi.org/10.1145/1235

Tian, H., & Zhuo, H. H. (2017). Paper2vec: Citation-Context Based Document Distributed Representation for Scholar Recommendation. *arXiv Preprint arXiv:1703.06587*.